%% file: impol-f.tex
\documentstyle{laa-s}  
\begin{document}
\thesaurus{03(03.09.5; 03.20.5; 02.16.2)}
\title{An Imaging Polarimeter(IMPOL) for multi-wavelength observations}  
\author{A. N. Ramaprakash\inst{1}\and  Ranjan Gupta\inst{1} \and 
A. K. Sen\inst{2}\and S. N. Tandon\inst{1}}
   \offprints{A. N. Ramaprakash}
 \institute{      
      Inter-University Centre for Astronomy and Astrophysics, Post Bag 4, 
      Ganeshkhind, Pune - 411 007, INDIA
      \and
      Center for Plasma Physics, Sapta Shwahid Marg, Dispur, 
      Guwahati - 781 006, INDIA
             }
\date{Received; Accepted}
%
%\maketitle
\markboth{A. N. Ramaprakash et al.: An Imaging Polarimeter (IMPOL)}
{A. N. Ramaprakash et al.: An Imaging Polarimeter (IMPOL)}
\abstract{
Taking advantage of the advances in array detector technology,
an imaging polarimeter (IMPOL) has been constructed for measuring linear 
polarization in the wavelength band from 400-800~nm. It makes 
use of a Wollaston prism as the analyser to measure
simultaneously the two orthogonal polarization components that define a 
Stoke's parameter. An achromatic half-wave plate is used to rotate the 
plane of polarization with respect to the axis of the analyser so that the 
second Stoke's parameter also can be determined. With a field of view correponding to about $\rm 30\times30~mm^2$ for a \diameter 1.2~m, f/13
telescope, a sensitive, liquid-$\rm N_2$ cooled CCD camera as the
detector and a built-in acquisition and guidance unit, the instrument can be used for studying stellar fields or extended objects with an angular 
resolution of $\sim $2\arcsec\ . The instrumental polarization is less than 
0.05\% and the accuracies of measurement are primarily limited by photon 
noise for typical observations.} 
\maketitle
\input psbox.tex
%

\section{Introduction}

The advances in two dimensional array detector technology in the 
optical and near infrared wavelength bands have made new kinds 
of imaging astronomical observations feasible.  Astronomical polarimetry 
is one field which has 
gained tremendously from these developments. The limitations of using
aperture photometry for polarimetry were so severe that any serious study 
was rendered time consuming and difficult. On the other hand, imaging
polarimetry with its capabilities for multiplexing,
simultaneous sky measurement, seeing-limited resolution etc. offer great
advantages over aperture polarimetry. Astronomers have recognized this 
potential and have developed several new observation techniques in the optical
(eg. Scarrott 1991; Jannuzi et. al. 1993; Jarrett et. al. 1994; Wolstencroft
et. al. 1995; Simmons~et.~al. 1995) and near infrared (eg. Kastner \& Weintraub
1994; Moore \& Yamashita 1995; Weintraub et. al. 1995) wavelenghts to study 
phenomena in a variety of Galactic and extragalactic astrophysical objects. 
In this paper, we report the design and construction of an imaging polarimeter
(IMPOL) which uses a cooled CCD array as detector. It was developed at the
Inter-University Centre for Astronomy and Astrophysics \hbox{(IUCAA)}, INDIA. 
The principle of the instrument (Sen \& Tandon 1994) is based on a combination
of ideas suggested
by Ohman (1939) and Appenzeller (1967). An instrument of this type has 
been constructed at the University of Durham and 
has been in use for some time now (Scarrott~et. al. 1983). Section~2 is 
a description of the instrument -- the different subsections dealing 
with various aspects ranging from design guidelines to instrument control 
and user-interface. The dominant sources of errors in the measurement are 
investigated in Sect.~3, while an estimate of the performance of the 
instrument under two typical observing conditions is given in Sect.~4; 
Section~4 also contains an estimate of the performance of the acquisition 
\& guidance unit. In Sect.~5 we discuss the observational procedure and 
the data-analysis method. Section~6 contains results of the commissioning
tests of the instrument. The last section (Sect. 7) contains the 
concluding remarks.

\section{The Instrument}

The following set of scientific and technical guidelines were adopted as
basis for the design of the instrument. (i)~It should be
capable of observing faint extended objects like reflection nebulae, 
accretion disks, dusty active galaxies etc., with
accuracy limited by photon noise and resolution limited by
seeing. This demands a typical field of view of a few arcminutes, an optics 
which is well-matched with the telescope to minimize loss of light 
and a detector
of sufficient sensitivity. Also, the effects of atmospheric scintillation 
should be eliminated and instrumental polarization minimized so that such
measurements of low level polarization could be made. (ii)~Since the objects 
of interest are very often quite faint, the instrument should have an 
acquisition and guidance (A\&G) system which allows pointing the telescope 
with an accuracy of a few arcseconds and tracking to better than 1\arcsec\  
over long periods. 
(iii)~Multi-wavelength observations should be possible in various wavelength
bands in the optical and near-IR regions. (iv)~The entire instrument 
consisitng of the optics, A\& G unit, associated
electronics etc. should be a self-contained unit which can be easily mounted 
on the telescope, with only electrically isolated communication links to 
the computers for instrument control and data acquisition. (v)~The cost
of the instrument should be minimized by using standard optical and
electronic components and by avoiding over-specification as far as possible. 

\subsection{Principle of the instrument}

Figure 1 is a schematic representaion of the optical arrangement of IMPOL. 
The basic idea behind this arrangement is to use a Wollaston prism with its 
axis normal to the optical axis of the system, as the analyser to convert 
the linear polarization in the incoming light into relative intensity of 
two orthogonally polarized beams (the ordinary and the extraordinary), 
separated by a small angle of 0.5\degr.
This measurement is sufficient to define one of the Stoke's parameters 
Q or U. The other Stoke's parameter can be determined by rotating the 
plane of polarization of the incoming light relative to the analyser
through a known angle and taking another 
measurement. This is accomplished by introducing a half-wave plate, with its 
fast-axis normal to the optical axis of the system, before the 
Wollaston prism. When the half-wave plate is rotated through an angle $\alpha$, 
the plane of polarization rotates through an angle $2\alpha$. At this new
position of the half-wave plate another measurement on the orthogonally
polarized beams can be made to determine the second Stoke's parameter as well.
It is easily seen that, for this arrangement the intensities 
of the extraordinary and ordinary beams ($I_{\rm e} \ and \ I_{\rm o}$) 
are given by

\vskip 2pc

\pscaption{\psboxto(\hsize;0cm){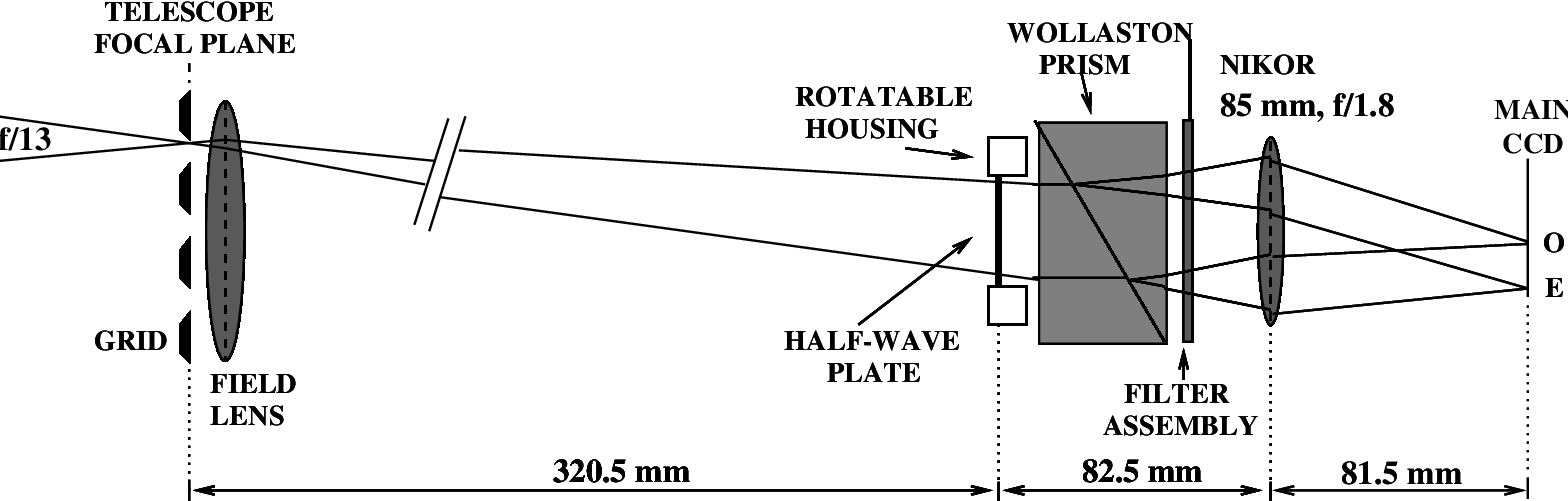}}{Fig. 1 Schematic of the 
IMPOL optical layout}
\medskip
%
%\begin{figure*}
%\picplace{6cm}
%\caption{Schematic of the IMPOL optical layout}
%\end{figure*}
%

\begin{eqnarray} 
I_{\rm e}(\alpha) &= &{I_{\rm up} \over {2}} + I_{\rm p} 
\cos^2(\theta - 2\alpha) \\ 
I_{\rm o} (\alpha) &= &{I_{\rm up} \over {2}} + I_{\rm p} \sin^2(\theta -
2\alpha)\;,
\end{eqnarray}
\noindent where $I_{\rm up} \ {\rm \&} \ I_{\rm p}$ are the intensities 
in unpolarized and polarized condition respectively in the incoming beam; 
$\theta \ {\rm\&} \ \alpha$ are the position angles of the polarization
vector and the half-wave plate fast-axis respectively with reference to the 
axis of the Wollaston prism. Since the angle $\theta$ is conventionally 
measured with respect to the celestial north-south axis (and increasing 
counter-clockwise), the axis of the Wollaston prism is kept aligned to it.
We define the ratio
\begin{equation}
R(\alpha ) = {{I_{\rm e} \over I_{\rm o}} - 1 \over {{I_{\rm e} \over 
I_{\rm o}} + 1}} = p\cos(2\theta - 4\alpha)\;,
\end{equation}
\noindent where  $p = {I_{\rm p} \over {I_{\rm up} + I_{\rm p}}}$, is 
the fraction of the total light in linearly polarized condition. This ratio
reduces to the normalized Stoke's parameters ${Q \over I}$ and ${U \over I}$
for $\alpha = 0\degr$ and $\alpha = 22.5\degr$. In practice, additional
measurements are made at two more values of $\alpha$, namely 45\degr\ 
and 67.5\degr, for reasons explained in Sect.~5. 

The half-wave plate and the Wollaston prism are placed 
in between a field lens--camera lens combination, which reimages the 
telescope focal plane on to the main CCD with a reduction factor of about 2.7; 
the field lens reimages the telescope aperture on the half-wave plate and 
the light reaches the camera lens without any vignetting. As shown in Fig.~1
each point in the telescope focal plane produces two images on the CCD,
corresponding to the ordinary and the extraordinary beams. In order to
avoid overlap of the images of adjacent points, for observations of 
extended objects, a grid of parallel obscuring strips is placed at the 
focal plane of the telescope. The width of and spacing between 
the strips are chosen in such a way as to avoid the overlap of the 
ordinary and the extraordinary images on the CCD. Four 0.3~mm diameter holes 
are provided at the four corners of the grid and are used to focus
the surface of the grid, which coincides with the focal plane of the telescope,
on to the surface of the CCD. The grid is 
made of black dielectric material to avoid polarization of the stray light
arising due to reflections from its edges. Besides, the edges are made slanted
(Fig. 1) to prevent vignetting of the telscope beam. 

\subsection{Acquisition and Guidance Unit}

A schematic illustration of the Acquisition and Guidance (A\&G) unit
is shown in Fig. 2. The right-angle prism, the field lens (L1) and a 
relay lens (L2) are held in position inside a probe which in turn is 
mounted on a linear XY translator stage. This arrangement allows the probe to 
be positioned anywhere within a $\rm 40\times 100~mm^2$ rectangular area
(corresponding to about 45 square minutes of arc for 

\vskip 2pc

\pscaption{\psboxto(\hsize;0cm){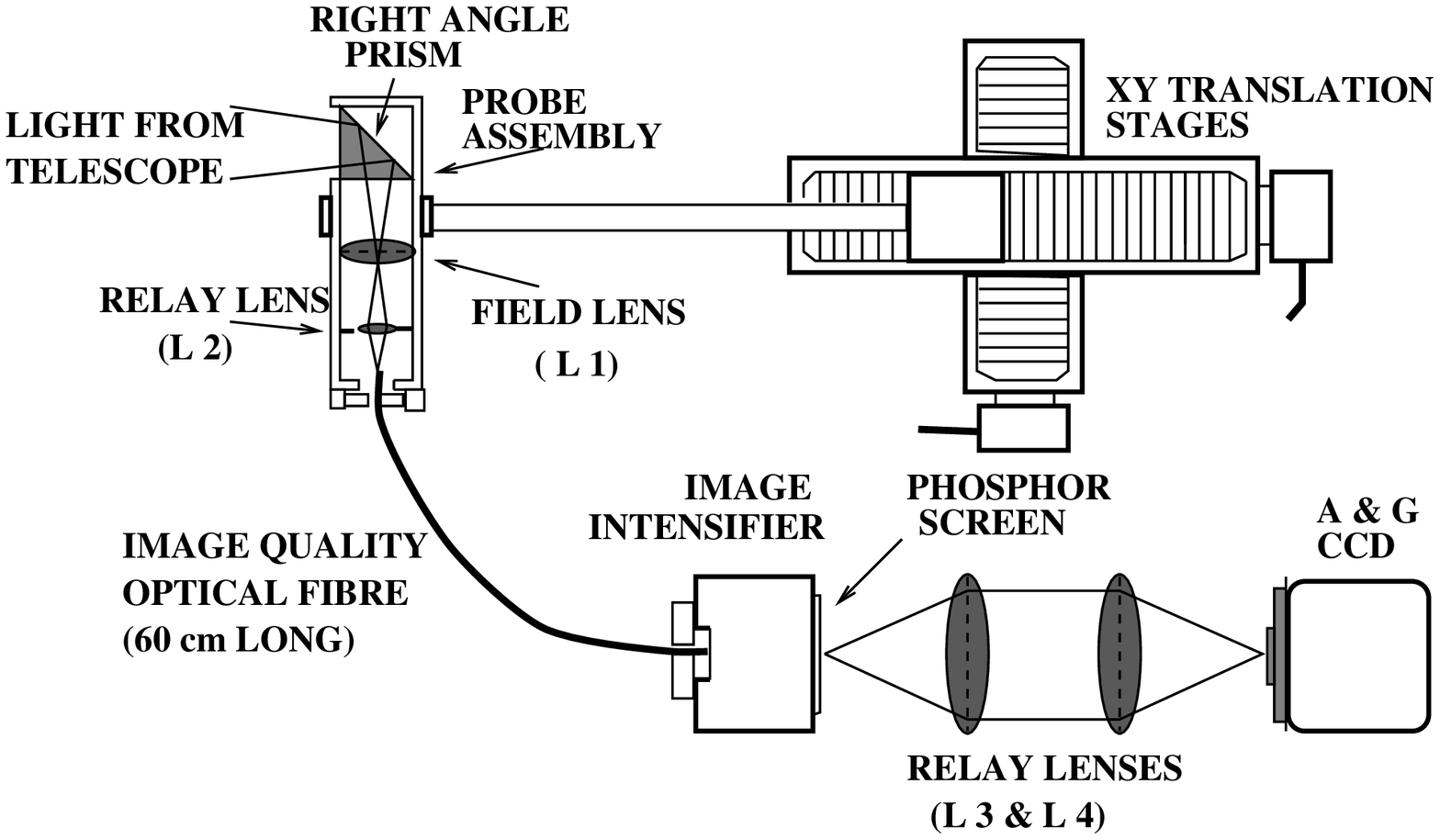}}{Fig. 2 Schematic of the 
Acquisition \& Guidance unit}
\medskip
\medskip
%
%\begin{figure*}
%\picplace{6cm}
%\aption{Schematic of the Acquisition \& Guidance unit}
%\end{figure*}
% 
\noindent a \diameter 1.2~m, f/13
telescope) close to the main field. The lens L2 reimages a \diameter 12~mm
area of the telescope focal plane on to the tip of an image quality 
optical fibre which, in turn, conveys it to the photocathode of an 
image intensifier. Two Nikor lenses (L3 and L4) focus the intensifier output 
on the surface of the CCD of the A\&G unit.

For pointing the telescope, atleast one star brighter than about $15^{\rm th}$
mag. in visual and close to the main field is identified and its right 
ascension and declination values are entered in to the control computer 
along with those of the main object. Once the translation stages are moved
to the requisite offset coordinates by the computer, short exposures are 
taken with the CCD of the A\&G unit to adjust the telescope pointing such 
that the guide star appears at the center of the CCD. Now the telescope has 
been pointed and auto-guiding is started. During auto-guiding the control
computer takes successive exposures with the CCD and finds the centroid
of the guide star image. Error signals are then generated
accordingly and transmitted to the control system of the telescope.
\begin{table*} 
\caption{Optical components used in IMPOL}
\begin{flushleft}
\begin{tabular}{lccr}
\hline\noalign{\smallskip}
Component & Source & Part number & \multicolumn{1}{c}{Description} \\
\noalign{\smallskip}
\hline\noalign{\smallskip}
Field lens & Karl Lambrecht & 322305 & \diameter 50~mm, eff. FL 300~mm, \\
& & & AR coated, achromat \\
Half-wave plate & Karl Lambrecht & WPAC 2-22-BB400/700 & achromatic, 
22~mm clear aperture, \\
& & & broad band single layer AR coated \\
Wollaston prism & Bernard Halle & PWQ 30.30 & quartz, 29x29~mm clear 
aperture, \\
& & & multi layer AR coated \\
Filters & Andover & & B, V and R, \diameter 40~mm, 5~mm thick, \\
& & & AR coated, image quality \\
Camera lens & Nikon & & eff. FL 85~mm, f/1.8, A.F. \\
\noalign{\smallskip}
Right-angle prism & Melles Griot & & 12x12~mm \\
\noalign{\smallskip}
Lenses L1 & Edmund Scientific & & eff. FL 25~mm, f/2.0 \\
\hskip 3.2em L2 & & & eff. FL 6~mm, f/2.0 \\
\noalign{\smallskip}
Optical fibre & Dolan Jenner & ISO 636 & 6x6~mm, coherent glass fibre, \\
& & & 36 inch long, N.A. 0.66, \\
Image intensifier & Philips & XX1500 & \diameter 14~mm, resolution 36~lp/mm \\
\noalign{\smallskip}
Lenses L3 \& L4 & Nikon & & eff. FL 50~mm, f/1.2 \\
\noalign{\smallskip}\hline
\end{tabular}
\end{flushleft}
\end{table*}

Table~1 is a list of the optical components used in IMPOL along with their 
sources and part numbers. The field lens, half-wave plate, Wollaston prism
and the filters were unmounted when purchased.

\subsection{Instrument control, Data-acquisition and User-interface}

A mixture of digital and analogue electronics hardware and a layered
architecture software have been used to perform the instrument control 
and data-acquisition operations. Apart from the control computer (PC 486), 
all the electronics is mounted on the telescope along with the instrument. 
The upper three layers of the software reside in the control computer, while
the lower two layers are downloaded into the local memory of the instrument 
at the time of start-up and initialization.

\medskip

\pscaption{\psboxto(8.7cm;0cm){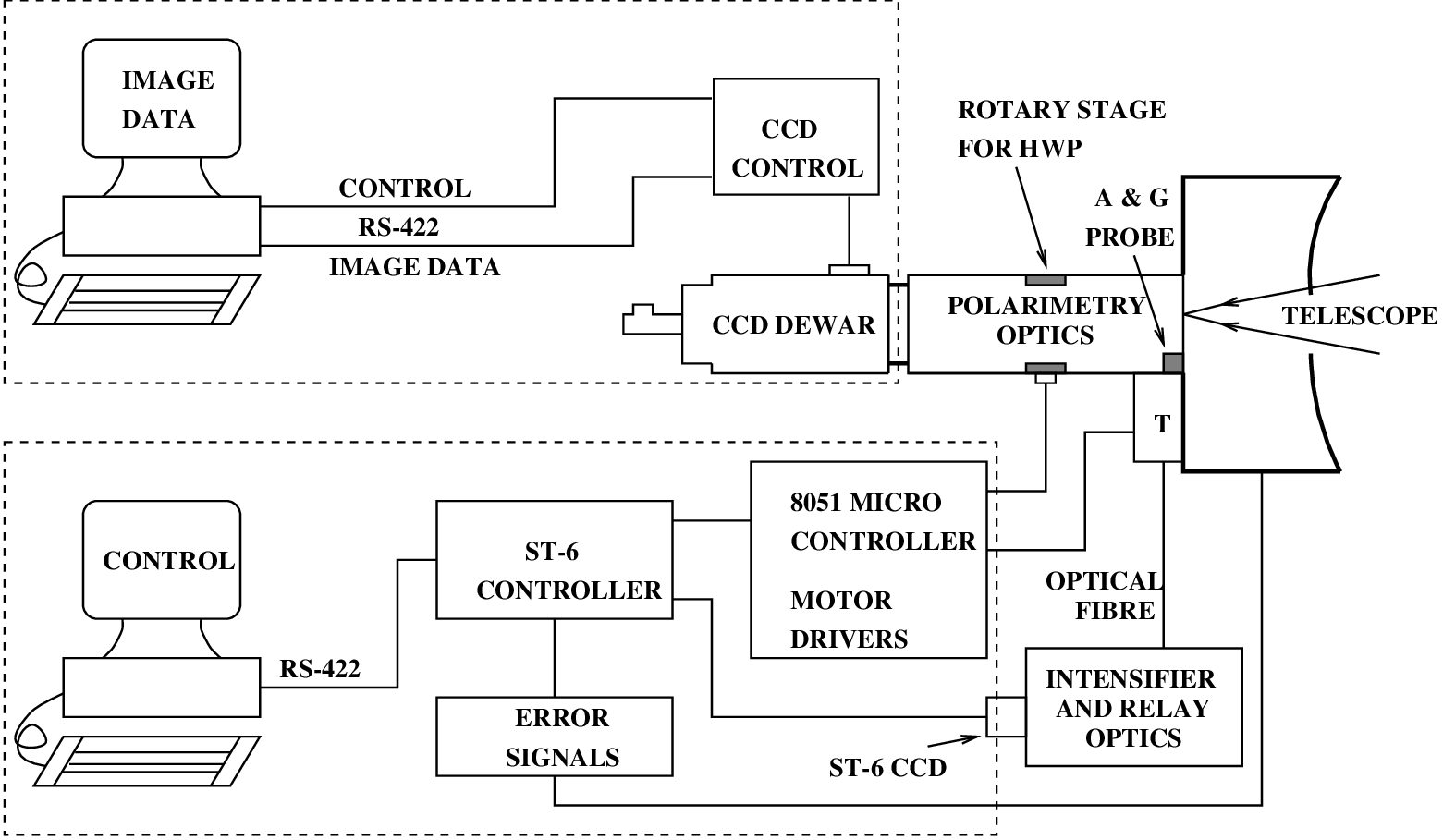}}{Fig. 3 Various blocks of the IMPOL
control system. The block marked ``T" represents the XY stages for positioning
the probes.}

\medskip 

%
%\begin{figure*}
%\picplace{7cm}
%\caption{Various blocks of the IMPOL control system. The block marked ``T"
%represents the XY stages for positioning the probe.}
%\end{figure*}    
%
\subsubsection{Electronics}

The electronics assembly of the instrument consists of three parts 
- (i) the positioning systems for the half-wave plate and the A\&G unit 
probe, (ii) the CCD camera for the A\&G unit and the associated electronics 
and (iii) the main CCD camera with its own control electronics and host 
computer (PC 486) for exposure control and data-acquisition. A block 
schematic of the IMPOL control system is shown in Fig.~3. 
The half-wave plate and the A\&G unit probe are positioned using stepper
motors which are controlled with signals generated by a 8051 microcontroller
based card. These digital control signals are fed to 
translator circiuts which convert them to power-amplified phase signals 
required to drive the stepper motors. Sensors are used on the half-wave
plate mount and the A\&G unit XY positioning stages for fixing the zero
posotions. The controller card, 
in turn, receives ``command" packets from the computer for instrument control.
 
The detector for the A\&G unit is the commercially available ST-6 CCD camera
of the Santa Barbara Instrument Group (SBIG). This camera has a controller
unit of its own which executes specific tasks under the command of the
control computer. (The SBIG have, in a private communication, graciously 
provided us the instruction set for their controller which we have used in our custom-software for instrument control.) The ``command-response" exchange 
between the control computer and the instrument takes place over an 
optically isolated asynchronous serial communication link (RS 422 standard).
The ST-6 controller has a second serial communication port (AUX port) so that 
it can be used as a gateway for communicating with other peripherals which use
the same type of communication link. The 8051-based controller card, described above, is connected to this AUX port and the communication between it and 
the control computer is routed through the ST-6 controller (see Fig.~3). 

The main CCD camera for data-acquisition is built around a liquid N$_2$ 
cooled, 385$\times$578 pixel, front-side illuminated, phosphor coated CCD 
chip from EEV. The operations of the CCD camera like shutter control, 
binning mode selection etc. are carried 
out by another 8051-based controller card which communicates with the 
data-acquisition computer over a second asynchronous serial communication 
link. Once the exposure is over the image pixal data are transmitted to the 
computer over a high speed cable using dedicated circuits. The important parameters of this CCD camera, which was also developed at IUCAA, are listed 
in Table 2 (for more details refer to Deshpande and Gadre 1994).
\begin{table} 
\caption{The parameters of IUCAA CCD Camera}
\begin{flushleft}
\begin{tabular}{ll}
\hline\noalign{\smallskip}
Parameters & Value \\
\noalign{\smallskip}
\hline\noalign{\smallskip}
CCD make & EEV CCD02-06 series \\
CCD chip size & 385(H)$\times$578(V) \\
Pixel size & $22 \times 22~{\rm \mu m}$ \\
Active area & $12.7~{\rm mm} \times 8.5~{\rm mm}$ \\
No. of amplifiers & 1 \\
Quantum efficiency & blue : 20\% \hfil \\
& yellow : 50\% \hfil \\
& red : 60\% \hfil \\
Readout speed & $32~{\rm \mu s}$ per pixel \\
Acquisition \& Display time & 12~s for full frame \\
Read noise (total) & 8~${\rm e^-}$ rms \\
Gain & 5~${\rm e^-}$ / ADU \\
\noalign{\smallskip}
\hline
\end{tabular}
\end{flushleft}
\end{table}

\subsubsection{Software}

The instrument control software is made of five layers, of which none but 
the top most one is visible to the user. This layer provides the
user-interface, reads the system configuration files and initializes the 
system. It is at this level where the sequence of operations is decided,
depending upon the instructions given by the user and corresponding service
calls to the immediate lower level are generated. The second layer provides
these services mostly by creating a ``command+parameter(s)" packet depending 
on the service requested and the current system configuration and then passing 
this packet on to the third software layer. The communication protocol 
between the control computer and the microcontrollers at the instrument 
is defined in the third layer which sents the packet to the ST-6 controller 
or through it to the 8051-based controller for positioning operations. 
Upon receiving a ``command+parameter" packet, either of these controllers 
parses it and checks for errors. If the command is valid and there are no 
errors it immediately sends back an acknowledgement signal. Otherwise it
sends back a corresponding error signal. The result of this communication
process is conveyed back to the highest layer for appropriate follow up. 
It is to be noted that at this stage the top most layer does not have 
information whether its service request has actually produced the required 
effect or not. To obtain this information it has to generate another service
call requesting the current status of the appropriate sytem parameters. 

The 8051 microcontroller has a built-in interpreter for BASIC. The 
processes of packet reception, parsing, error checking and acknowledging 
are carried out by a BASIC program which resides in the memory of 
the microcontroller forming the fourth software layer. A set of assembler
routines, each doing a specific operation, forms the fifth and lowermost
software layer. The BASIC program calls one of these assembler routines
to execute the particular task requested by the command. 

This system of control encashes on the fact that while each individual 
operation is to be carried out fast, the interval between successive 
operations are comparable to typical user response time scales. Though 
the layered architecture appears complicated, it is in fact extremely
easy to implement and debug and provides great flexibility to alterations.

\section{Sources of errors}

In addition to the fundamental limit due to photon statisitcs, the measurement
of polarization is affected by several factors beginning from the atmosphere
down to the detector and the data-analysis procedure. In the following four
subsections we look at, in some detail, the important sources of errors 
and the techniques used to minimize their effects. 

\subsection{Atmospheric Effects}

Since polarization measurement involves taking multiple exposures,  
variations in atmosperic transparancy and scintillations can affect 
the measurement process.
In IMPOL, two orthogonal polarization components are measured simultaneously 
and the Stoke's parameter is obtained from the ratio of the fluxes in these
components (Eq. (2)). Since the atmosphere 
is not birefringent, this eliminates the effects due to atmospheric
scintillation, or for that matter any effect which changes both the 
polarization components by the same factor--like variations in the effective
exposure times of observations, presence of thin clouds etc. 

\subsection{Distortions in polarization}

As the measurements depend on detecting changes in the polarization
vector caused by rotating the half-wave plate, any small instrumental
polarization occuring after the half-wave plate can be easily eliminated. 
All the optical elements are anti-reflection coated to minimize the 
polarization efffects due to reflection at their surfaces, and care is taken 
to minimize the stray light, reflected from the walls etc., reaching the 
detector. In particular, all the aperture stops in the beam path are chosen
to be non-metallic, including the grid in the focal plane, so as to avoid
polarization of light scattered from these (Pospergelis, 1965). The mounts for
field lens, the half-wave plate and the Wollaston prism have been designed
to minimize stress-birefringence due to differential thermal expansion. 

The retardance introduced by the half-wave plate could deviate from 180\degr\
either because of the finite angle of incidence or because of the chromatic
effects. If the beam is incident at a small angle $i$, the 
maximum change in retardance is given by (derived from expression in 
Serkowski 1975)
\begin{equation}
\sigma_\tau \simeq \pi\times {i^2 \over
{4n_{\rm o}}}\left ({1 \over {n_{\rm o}}} - {1 \over {n_{\rm e}}}\right )\;,
\end{equation}
where $n_{\rm o}$ and $n_{\rm e}$ are the refractive indices of the material
for ordinary and extraordinary rays. For the aperture used, the maximum 
angle of incidence is about 5\degr\ so that $\sigma_\tau \la 0.01$~rad. 
The chromatic effects give $\sigma_\tau < 0.18$~rad for the wideband and 
$\sigma_\tau < 0.14$~rad for the V-band. It can be shown (Serkowski 1974) 
that the depolarization $\sigma_p$ due to an uncertainty of $\sigma_\tau$~rad 
in the retardance is to the lowest order given by 
$\sigma_p \simeq p\sigma_\tau^{\rm 2}$, $p$ being the fractional polarization.
Therefore even for the wideband $\sigma_p \la 0.03p$. Further, circular
polarization in the incident 
light is converted to a linear polarization of magnitude $V\sigma_\tau$, where 
$V$ is the circularity parameter. For wideband observations, conversion to 
linear polarization would be about $0.18V$, but in typical observations this 
does not pose a serious problem because the circular polarization in the
incident light is usually less than the linear polarization.  

Another chromatic effect is the change in position angle of the 
half-wave plate
fast axis with wavelength. Although this does not produce any error in the
measurement of fractional polarization $p$, it does render the measurement
of position angle~$\theta$ erroneous. A solution to this problem is to  
observe standard stars using narrow band filters and produce a smoothed curve
which gives the required position angle correction for each wavelength. 
The achromatic half-wave plate used in IMPOL does not produce any appreciable
dispersion in its fast-axis position angle over the wavelength range of
interest. 

The uncertainty involved in the positioning of the half-wave plate
leads to an error in the measurement of polarization. From Eq. (2), we can 
see that a positioning uncertainty of $\sigma_\alpha$ produces a maximum 
error in the measurement of linear polarization given by $\sigma_p \simeq 
p\sigma_\alpha$. Thus for an error of 0.1\degr, of the postioning system, 
this results in $\sigma_p \simeq 0.002p$.   

\subsection{Flat field errors}

The half-wave plate is the only moving component in the optical train and
hence it is relatively easy to make sure that the images do not shift on 
the detector. Further, as the analyser is fixed with respect to the
detector, the orientation of the ordinary and extraordinary polarizations 
also remain fixed with respect to it, independent of the polarization vector 
of the incident beam. This renders it easy to estimate the 
flat field correction factor from the set of observations at the four
positions of the half-wave plate. These points are further explained in Sec. 5.

\subsection{Photon noise}
 
From the above discussion we find that all the sources of errors described 
do not contribute more than 0.05$p$ to the error in the measurement 
of $p$. In the next section, we estimate the errors under typical observing
conditions that result from photon noise alone. Comparing the two, we conclude
that the dominant source of error in the typical measurement of polarization 
is photon noise. From equation (2) it can be shown that the variance due to 
photon noise in the measurement of each of the Stoke's parameters $R(\alpha)$
is given by  
\begin{equation}
\sigma_R^2 = {4I_{\rm o} I_{\rm e} \over {(I_{\rm o} + I_{\rm e})^3}}(1 + k)
\;,
\end{equation}
where $I_o$ and $I_e$ are the number of photoelectrons and $k$ is the ratio 
of the flux from the background to that from the source. For small values 
of $p$, $I_{\rm o}$ and $I_{\rm e}$ are approximately
equal and denoting $I = I_o + I_e$, the standard deviation of $p$ can be 
written as   
\begin{equation}
\sigma_p \simeq 100 \times {\sqrt{I + I_{\rm B}} \over {I}}\;\%\;,
\end{equation}
where $I_{\rm B} = kI$. Similarly, we can also show that 
\begin{equation}
\sigma_\theta \simeq 0.5 \times {\sigma_p \over p}~{\rm rad}\;.
\end{equation}

It is worth noting here that $p$, as defined above, is a positive definite 
quantity and follows the Rice probability distribution given by
\begin{equation}
F(p, p_0) = {p \over \sigma_p}e^{p^2 + p^2_0 \over 2\sigma_p^2} 
I_{0}({pp_0 \over \sigma_p^2})\;.
\end{equation} 
Here, $p_0$ is the true value of fractional polarization being estimated by
$p$ and $I_0$ is the modified Bessel function of order zero. Since $p$ is 
a biased estimator, several schemes have been suggested
(Simmons \& Stewart 1984 and references therein) for debiasing, but none of 
them is fully satisfactory. For values of ${p \over \sigma_p}$ larger than
about 4, almost all the debiasing schemes agree and reduce approximately to 
the relation
\begin{equation}
\hat p_0 = p \sqrt{ 1- {\sigma_p^2 \over p^2}}\;, 
\end{equation}
where $\hat p_0$ is the ``debiased" estimate of $p_0$. Thus it is desirable to 
work with the normalized Stoke's parameters as far as possible 
and use the quantities $p$ and $\theta$ only to present the final results. 
However, in general, the normalized Stoke's parameters themselves might not be
normally distributed and might have bias (Clarke et. al. 1983). Besides, for
photon-noise dominated measurements, the positive kurtosis of the distribution
will lead to erroneous estimates of the confidence levels unless more than 
a few thousand photoelectrons are collected. Thus, in order to arrive at an 
optimum procedure, it is essential to carefully study the nature of the 
dominant sources of noise in the measurement and their effects on the
analysis (see Sec.~5). 

\section{Performance estimates} 

An estimate of the performance of the instrument is made in the following  
two sub-sections. In the first part we calculate the accuracy of 
polarization measurement as limited by photon noise for two typical types of
observations. The second part deals with the performance of the A\&G unit.
In both cases, it is assumed that the observations are made in the V-band
(bandwidth $\Delta\lambda\sim~100~{\rm nm}, V_0~=~1000$~photons 
per $\rm cm^2$ per s per $\AA$) using a \diameter 1.2~m 
(area $A \sim 10^4~{\rm cm^2}$) telescope having an f/13 beam, 
that the brightness of the sky is 20~mag. per sq. arcsec and that the
software aperture used for photometry has about 30~pixels, ie. about 
30~sq. arcsec.

\subsection{Polarimetry}

In the following discussion the combined effect of the atmosphere, the
telescope, the optics and quantum efficiency of the detector etc., is found 
to result in an effective transmission of $\eta \sim 3.2\%$.

A classical example of polarimetric observations involves study of dark
molecular clouds with the light of stars shining behind their periphery
(eg. Elvius 1970; Joshi et al. 1985; Kane et al. 1995).
Such studies give valuable information regarding the properties of  
dust in the cloud and the magnetic field in its vicinity. Typically the
polarization of these sources is in the range \hbox{1--5~\%.} Assuming stars
of apparant visual magnitude $m_{\rm v} = 15$, we see that the total number 
of photo-electrons collected, due to the star ($I$) and the background 
($I_B$), within the software aperture, for an exposure of 15~minutes are 
$2.9 \times 10^5$ and $8.6 \times 10^4$ respectively. Therefore, the corresponding error in polarization measurement is given by
\begin{equation}
\sigma_p = 100 \times {\sqrt{2.9 \times 10^5 + 
8.6 \times 10^4} \over {2.9 \times 10^5}} \simeq 0.21\%\;.
\end{equation}
If the observations are made with a filter which covers both
the V and R bands ($\Delta\lambda\sim 200~{\rm nm}$), there will be 
an improvement in the signal-to-noise ratio by a factor of $\sqrt{2}$, giving
$\sigma_p\sim 0.15\%$.

Study of extended objects like reflection nebulae, is another field in 
which polarimetric observations are useful. Polarization in the light 
scattered from these clouds can range from a few percent to as much 30\%.
Assuming that the observation involves a reflection nebula of the same 
surface brightness as the background sky, the integrated photoelectron count
from each, for an exposure of 15~minutes, will be $8.6 \times 10^4$ and hence 
\begin{equation}
\sigma_p  = 100 \times {\sqrt{8.6 \times 10^4 + 
8.6 \times 10^4} \over {8.6 \times 10^4}} \simeq 0.48\%\;. 
\end{equation}
%
%\begin{figure}
%\picplace{7.5 cm}
%\caption{The errors due to photon noise alone, as calculated in Sec. 4.1, 
%for wideband measurements of fractional polarization $p$ are shown 
%as a function 
%of the brightness of the source. The solid curve refers to the case of stars
%while the dashed curve is for an extended object, for which the x-axis
%represents the surface brightness of the source in magnitudes per sq. arcsecs.
%The background is assumed to be $20^{th}$ magnitude per sq. arcsec and a 
%software aperture of 30 sq. arcsec has been used during data reduction. 
%The crosses 
%represent the errors in actual polarimetric measurements made with instrument
%on a field at the periphery of the dark cloud B133.}
%\end{figure}
%
\medskip
\pscaption{\psboxto(\hsize;0cm){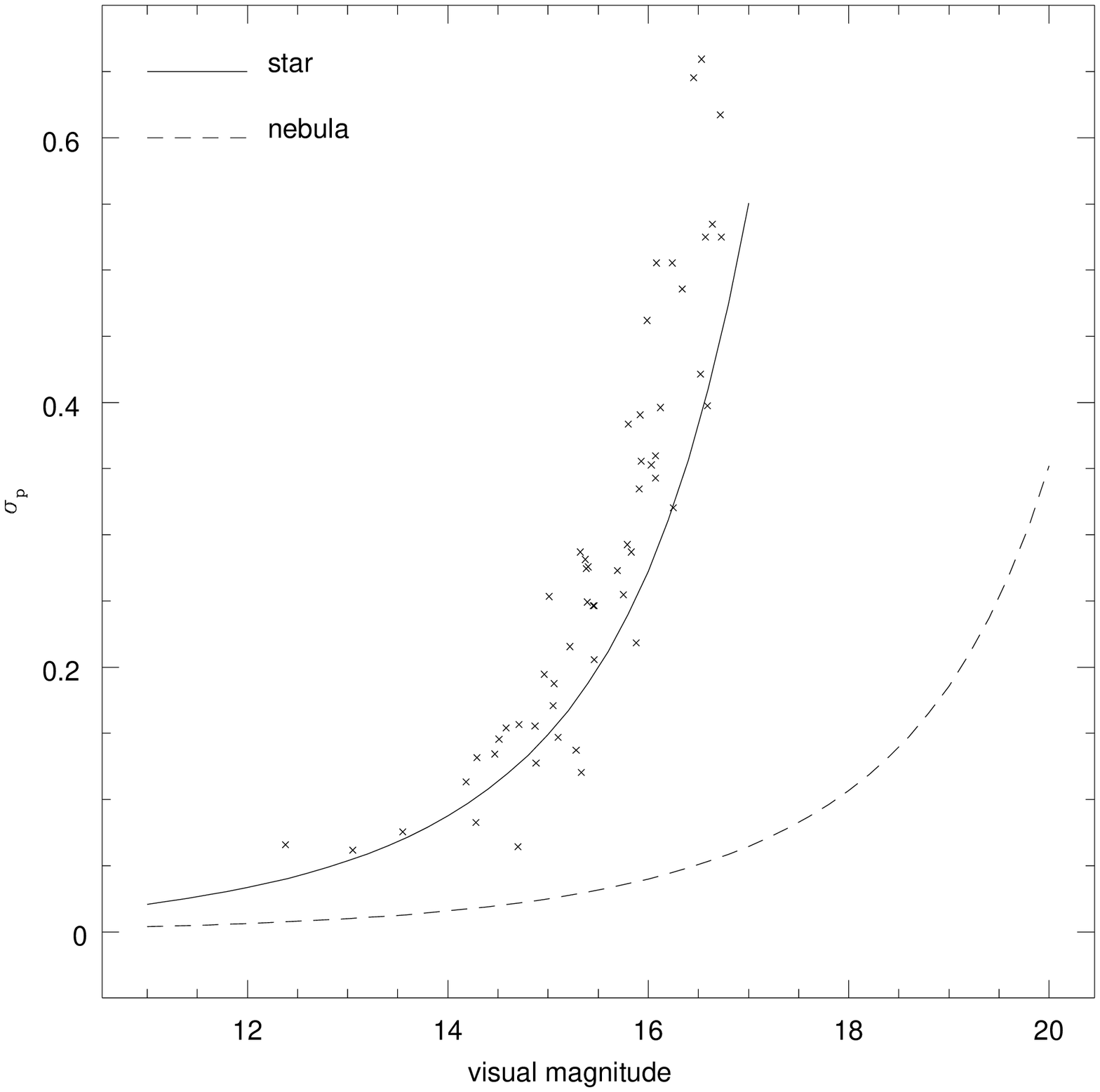}}{Fig. 4 The errors due to photon 
noise alone, as calculated in Sec. 4.1, for wideband measurements of 
fractional polarization $p$ are shown as a function 
of the brightness of the source. The solid curve refers to the case of stars
while the dashed curve is for an extended object, for which the x-axis
represents the surface brightness of the source in magnitudes per sq. arcsecs.
The background is assumed to be $20^{th}$ magnitude per sq. arcsec and a 
software aperture of 30 sq. arcsec has been used during data reduction. 
The crosses 
represent the errors in actual polarimetric measurements made with instrument
on a field at the periphery of the dark cloud B133.}
\medskip 

\noindent for a solid angle defined by the software aperture. The wideband 
observations, in this case will give an error of about 0.34\%. 

In Fig.~4 the solid and dashed curves show how
the error in the measurement of fractional polarization, based on photon
statistics alone, changes as a function of source brightness, for the two 
cases discussed above, namely stellar sources and reflection nebulae 
respectively. 

\subsection{Acquisition \& Guidance (A\&G) unit}

The A\&G unit has a field of view of about 2\arcmin\ and can be positioned 
anywhere within a 40$\times 100~{\rm mm}^2$ area at the focal plane of the 
telescope close to the main field. This area has been chosen on the basis of 
the requirement that three stars brighter than $m_{\rm v} = 15$ 
should be available with 95\% confidence level towards the poles of the 
Galaxy. 

In this case, apart from the atmospheric and telescope transmission losses, 
account has to be made for losses in the probe optics, optical fibre and
the quantum efficiency of the intensifier. Together this amounts to an 
effective transmission of $\sim$0.1. The gain of the intensifier can be varied
from 2000 to 50000 and the effective transmission of the lenses L3 and L4 and
the ST-6 CCD is about 3\%. Therefore the signal-to-noise ratio remains 
essentially unchanged in the stages after the intensifier. 
%
%\begin{equation}
%{\rm SNR} = {10^{3 - 0.4m_{\rm v}{\rm (star)}} \times 
%\sqrt{A (\Delta\lambda) \eta_1\eta_2 t_{\rm exp}} \over
%\sqrt {(10^{3 - 0.4m_{\rm v}{\rm (star)}} + 10^{3 - 0.4m_{\rm v}{\rm (sky)}}) 
%(\eta_2 + 1)}}\;,  
%\end{equation}
%
An exposure of 1~s for a star of V $\sim$15 mag., gives a signal-to-noise 
ratio of about 20 in each of the 10 pixels covered by the image. This is
sufficient to obtain an auto-guidance accuracy better than 0.1\arcsec\ with 
a correction 
frequency of about a second. However, in order to avoid excessive jitter,
hysterisis has been introduced into the system so that the corrections 
will occur only if a minimum shift of 0.2\arcsec\ has been sensed in the 
stellar image centroids. 

\section{Observations and Data-analysis}

We have seen in Sect. 2.1 that observations for at least two positions 
of the half-wave plate are required to determine the three 
unknown parameters namely, the total
intensity ($I$), fraction of light in linearly polarized condition ($p$),
and position angle of the plane of polarization ($\theta$). 
However, the situation is not so simple in reality because~--~(i)~the responsivity of the system to the two orthogonal polarization 
components may not be the same, and (ii)~the responsivity of the CCD
is a function of the position on its surface. Due to these 
effects the signals which are actually measured in the two images 
($I_{\rm e}^\prime$ and $I_{\rm o}^\prime$) are given by   
\begin{eqnarray}
I_{\rm e}^\prime(\alpha) & = & I_{\rm e}(\alpha) F_{\rm e}(x,y) 
{\rm\quad and \quad} \nonumber \\
I_{\rm o}^\prime(\alpha) & = & I_{\rm o}(\alpha) F_{\rm o}(x,y)\;,
\end{eqnarray}
\noindent where $F_{\rm e}(x,y)$ and $F_{\rm o}(x,y)$ represent the
effects mentioned above; ($x,y$) being the coordinates on the array
surface. If it is ensured that the ordinary and extraordinary images 
do not move on the surface of the CCD during different 
exposures, the ratio of the factors $F_{\rm o}$ and $F_{\rm e}$ can be 
estimated as 
%
%\begin{equation}
%I_{\rm e}^\prime(\alpha) = I_{\rm e}(\alpha) F_{\rm e} {\rm\quad and \quad} 
%I_{\rm o}^\prime(\alpha) = I_{\rm o}(\alpha) F_{\rm o}\;.
%\end{equation}
%
%Now the quantity $F = {F_{\rm o} \over F_{\rm e}}$ can be estimated as 
%
\begin{equation}
{F_{\rm o} \over F_{\rm e}} = \big [{I_{\rm o}^\prime(0\degr ) \over 
I_{\rm e}^\prime(45\degr )} \times 
{I_{\rm o}^\prime(45\degr ) \over I_{\rm e}^\prime(0\degr )} \times
{I_{\rm o}^\prime(22.5\degr ) \over I_{\rm e}^\prime(67.5\degr )} \times
{I_{\rm o}^\prime(67.5\degr ) \over I_{\rm e}^\prime(22.5\degr )}
\big ]^{1 \over 4}\;,
\end{equation}
by making use of the fact that a rotation of the half-wave plate by 45\degr\  
simply leads to an interchage of the signals in the ordinary and extraordinary
images. Now the actual ratio of the fluxes in the two images may be recovered 
as
\begin{equation}
{I_{\rm e}(\alpha) \over I_{\rm o}(\alpha)} = {F_{\rm o} \over F_{\rm e}} 
\times {I_{\rm e}^\prime(\alpha) \over I_{\rm o}^\prime(\alpha)}\;. 
\end{equation}
This ratio is substituted in Eq. (2) and a cosine curve is fitted to the four
values of $R(\alpha)$ obtained so as to make the best estimates of $p$ and
$\theta$.

In practice, it might also become necessary to take several exposures at each 
half-wave plate position so as to collect sufficient number of photons to
get the required signal-to-noise ratio. This might also become necessary 
if there are stars in the field with very large magnitude differences. Then 
a set of short exposures have to be taken to avoid the bright stellar images
from saturating while a set of longer exposures will be required to collect
sufficient photons from the fainter stars. In order to ensure that the
normalized Stoke's parameters follow a normal distribution as close as
possible, the basic observational quantities, namely the fluxes measured at
each position of the half-wave plate, should be averaged together before taking 
their ratios to give the normalized Stoke's parameters. 
However, this method has its demerits too. Firstly, it leaves 
the non-linear $\chi^2$ fitting technique with only one degree of 
freedom, to estimate the two parameters $p$ and $\theta$ from the four 
values of $R(\alpha)$. Secondly, it does not allow the estimation of errors
in the measurement of the Stoke's parameters, since there are not enough 
degrees of freedom. Therefore, it appears optimum to use a fitting technique 
to estimate $p$ and $\theta$, from the values of $R(\alpha)$ obtained 
from individual exposures. 

Since a grid of parallel obscuring strips is placed at the telescope 
focal plane (Sect. 2.1), only less than half the field is seen by the
detector during any observation. Once a set of exposures has been taken
for all the four positions of the half-wave plate, the telescope orientation
is slightly changed so as to see the other part of the field which was
previously obscured by the grid strips. Alternatively, in the case of 
stellar fields with slowly changing (in intensity and polarization) 
background, the grid may be removed during observations, provided the field 
is not too crowded. 

The images are collected by the data-acquisition computer during 
observations and later transferred to cartridge tapes. The data analysis is carried out on UNIX-based workstations. A polarimetry package 
has been developed for this purpose within the IRAF environment using a mixture 
of standard IRAF tasks, custom-made CL scripts and FORTRAN routines. The 
images are first subjected to the preliminary steps of bias and dark 
removal, cosmic ray and bad pixel detection, and masking etc. PSF fitting 
tasks of the DAOPHOT package are then used to determine accurately the 
centroids of the stellar images. The intensity estimates are, however made 
using aperture photometry covering a diameter greater than 
$2\times$FWHM so as to integrate more than 90\% of the signal. (The FWHM
is kept greater than 3 pixels to minimize the effects of undersampling
and intrapixel variation of quantum efficiency.) 

\section{Commisioning of IMPOL}

The commisioning of the instrument involved two levels of tests. Firstly
laboratory tests were conducted on the optical bench to determine the 
various parameters like linearity, gain and readnoise of the detector, vignetting, depolarization, instrumental polarization etc. Some necessary
modifications were made in the optical system and the software in order 
to optimize the instrument performance. Measurements were made with a 3000~K
filament source in B, V, R
bands, and a wide band defined by a 3~mm thick KG3 glass. A 100\% polarized 
beam showed that the depolarization is about 0.4\% in the B and
V bands, less than 1\% in the R band, and less than 6\% in the broad band. 
There was no discernible rotation of the position angle between these bands. 
For R and the wideband, the accuracy of measurements
is limited by the performance of the polarizer used. In order to
estimate the instrumental polarization, a Lyot depolarizer was introduced 
in front of the artificial star and the polarization was measured. This 
experiment showed that the instrumental polarization is about 0.03\% in the 
B-band and less than 0.06\% with the wideband filter. 

\begin{table}
\caption[ ]{Observations of standard stars}
\begin{flushleft}
\begin{tabular}{llllllll}
\hline\noalign{\smallskip}
\hfil Star \hfil & \hfil $p$ \hfil & \hfil $\sigma_p$ \hfil 
& \hfil $\theta$ \hfil & \hfil $\sigma_\theta$ \hfil 
& \hfil $\sigma_{\rm phot}$ \hfil & \hfil $p_0$ \hfil & \hfil $\theta_0$ \hfil\\
\noalign{\smallskip}
\hfil (1) \hfil & \hfil (2) \hfil & \hfil (3) \hfil & \hfil (4) \hfil & 
\hfil (5) \hfil & \hfil (6) \hfil & \hfil (7) \hfil & (8) \hfil \\
\noalign{\smallskip}
\hfil HD \hfil & \hfil \% \hfil  & \hfil \% \hfil 
& \hfil \degr \hfil & \hfil \degr \hfil & \hfil \% \hfil 
& \hfil \% \hfil & \hfil 
\degr \hskip 1em \\
\noalign{\smallskip}
\hline\noalign{\smallskip}
43384 & 3.05 & 0.04 & 172 & 0.3 & 0.03 & 3.0 & 170 \\
154445 & 3.65 & 0.03 & 90.5 & 0.2 & 0.02 & 3.7 & 90 \\
\noalign{\smallskip}
102870 & 0.03 & 0.01 & 64.4 & 7.8 & 0.02 & 0.017 & 162 \\
\noalign{\smallskip}
39587 & 0.03 & 0.03 & 140 & 27.5 & 0.02 & 0.013 & 20 \\
& 0.05 & 0.02 & 159 & 12.7 & 0.02 & & \\
& 0.09 & 0.02 & 34 & 6.7 & 0.02 & & \\
& 0.05 & 0.02 & 167 & 11.3 & 0.02 & & \\
\noalign{\smallskip}
\hline
\end{tabular}
\end{flushleft}
\end{table}

The second stage of tests where conducted when the instrument saw the 
first light in February 1996, at the \diameter 1.2~m Gurusikhar 
Infra-Red Telescope facility 
on Mt.~Abu, Rajastan, operated by the Physical Reearch Laboratory at
Ahmedabad, India. Several standard polarized and unpolarized stars were 
observed at this time. Since the standard stars are too bright, having 
visual magnitudes in the range of 4 to 7, instrument was slightly defocussed 
to avoid saturation of pixels in reasonable exposure times.

Table 2 contains the results of the observations in V-band. Column~1 of 
the table gives the identification number of the star. Columns~2-5 contain 
the measured values and
$1\sigma$ errors of the fractional polarization $p$ and the position angle
$\theta$. Column~6 gives the error $\sigma_p$ based on
photon statistics alone, derived as illustrated in Sect.~3. Cols.~7 and~8
list the published values of $p$ and $\theta$. 

The first two stars in the table are polarized standards while the last two 
are unpolarized ones. Comparing the measured and published results for 
the polarized stars we see that there is no indication of depolarization.
Neither is there any rotation of the position angle beyond a couple of degrees.
From Col.~3 and Col.~6 for the unpolarized stars, we see that 
the error in the measured polarization is comparable with that expected
on the basis of photon statistics alone. In Fig. 5, the values of 
$q = R(0\degr)$ are plotted against $u = R(22.5\degr)$ for all the unpolarized
standards observed in the V-band. The extremely low correlation coefficient
between $q$ and $u$, is indicative of the low instrumental polarization.
The average values of $q$ and $u$, give a value of polarization
$p = 0.03\%$. These observations show that for accuracies $\sog$0.05\% 
in the measurement of polarization $p$, the
performance is still limited by photon noise and the instrument polarization
floor has not yet reached. The four measurements of star HD39587 are 
made at different corners of the CCD frame in order to verify 
the uniformity of response. 

The data which 
is required to incorporate the A\&G system was also collected during this
observation run. The A\&G system was succesfully installed during a second
observation run with the same telescope in late May, 1996. This time, several 
observations of stellar fields at the periphery of some dark molecular 
clouds were also made. The analysis of this data is in progress at present 
and will be reported separately. The accuracy of the polarization
measurement can be gauged from the results for some of these stars shown
in Fig.~4.

\section{Conclusion}

An Imaging Polarimeter (IMPOL) has been constructed which uses mostly 
standard optical and electrical components, but through careful design 
has been able to achieve 

%\begin{figure}
%\picplace{7.5 cm}
%\caption{Normalized Stoke's parameter $q$ is plotted against $u$ for a
%number of unpolarized standard stars. The correlation coefficient of the
%points is about 0.06 and the average values of $q$ and $u$ give a 
%value of $p = 0.03\%$.}
%\end{figure}
%
\medskip
\pscaption{\psboxto(\hsize;0cm){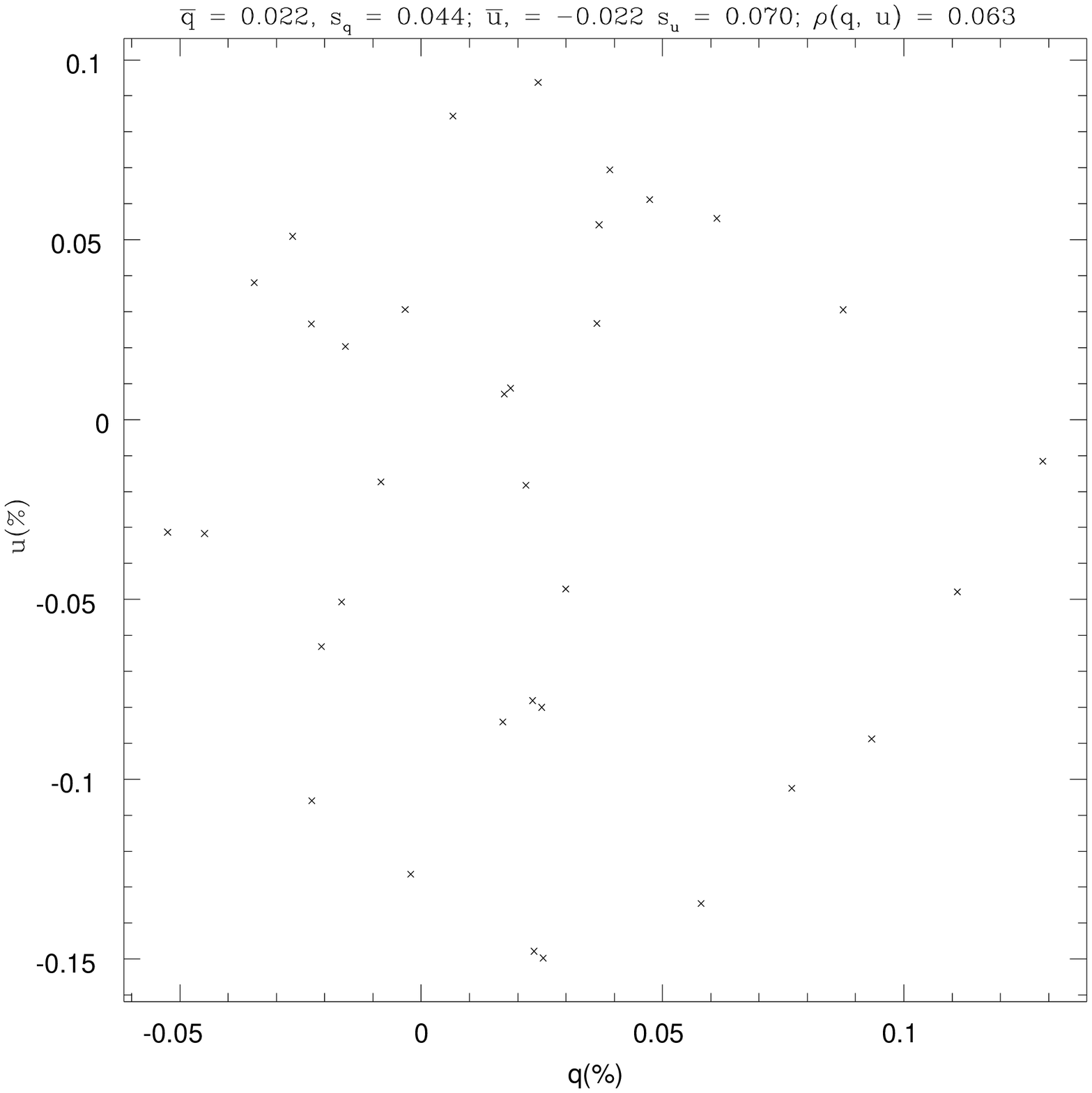}}{Fig. 5 Normalized Stoke's 
parameter $q$ is plotted against $u$ for a
number of unpolarized standard stars. The correlation coefficient of the
points is about 0.06 and the average values of $q$ and $u$ give a 
value of $p = 0.03\%$.}
\medskip 
\noindent photon-noise limited performance. The instrument has a field 
of view of about 6.5\arcmin\ for a \diameter 1.2~m, f/13 telescope, and the detector is a sensitive, liquid N$_2$ cooled CCD. An off-axis acquisition and guidance unit capable of using stars as faint as $V_{\rm mag}\sim 15$ is 
also built in the instrument so that long exposures of faint 
extended objects like reflection nebulae etc. can be taken while keeping
the image fixed on the CCD face to within one half of a pixel -- this stability
of the image is a necessary condition to achieve 
a high accuracy in relative photometry between the different 
frames used to estimate the polarization.
Observations of nearby standard polarized and unpolarized stars show that 
for wideband observations, there is no discernible depolarization and the
instrumental polarization is less than 0.05\%. Preliminary results of 
wideband polarimetry of stellar fields, using a PSF fitting technique for 
determining the centroids of the stellar images and aperture 
photometry with a radius of 3 pixels to derive the intensities of the 
individual images, give close to photon-noise limited  accuracy of 
0.15\% for $\rm V_{mag} = 15$ stars with about an hour of total exposure 
time. For extended objects with brightness about 20th mag. per square 
arcseconds and a background of
about the same brightness, it is estimated that an accuracy of about 1\%
is possible with 60 minutes of total exposure time and an aperture of
10~sq. arcsec.

\begin{acknowledgements}
The project has been funded by the Department of Science and 
Technology of the Government of India. The authors also wish to thank
the Physical Research Laboratory, Ahmedabad for providing telescope time,  
P.~A.~Chordia and R.~Bedade for their involvement in developing the 
electronic circuitry for the instrument and M.~S.~Deshpande for her 
assistance during its commissioning.
\end{acknowledgements}

\end{document}

%% file: psbox.tex
\def\temp{1.34}%
\let\tempp=\relax
\expandafter\ifx\csname psboxversion\endcsname\relax
  \message{PSBOX(\temp) loading}%
\else
    \ifdim\temp cm>\psboxversion cm
      \message{PSBOX(\temp) loading}%
    \else
      \message{PSBOX(\psboxversion) is already loaded: I won't load
        PSBOX(\temp)!}%
      \let\temp=\psboxversion
      \let\tempp= 
    \fi
\fi
\tempp
\let\psboxversion=\temp
\catcode`\@=11
% Every macro likes a little privacy...
%
%Trying to tame the variety of \special commands for Postscript: the
%  universal internal command \PSspeci@l##1##2 takes ##1 to be the
%  filename and ##2 to be the integer scale factor*1000 (as for usual
%   TeX \scale commands)
%
\def\psfortextures{%     For TeXtures on the Macintosh
%-----------------
\def\PSspeci@l##1##2{%
\special{illustration ##1\space scaled ##2}%
}}%
\def\psfordvitops{%      For the DVItoPS converter on IBM mainframes
%----------------
\def\PSspeci@l##1##2{%
\special{dvitops: import ##1\space \the\drawingwd \the\drawinght}%
}}%
\def\psfordvips{%      For DVIPS converter on VAX, UNIX and PC's
%--------------
\def\PSspeci@l##1##2{%
%    \special{/@scaleunit 1000 def}% never read dox without trying!
\d@my=0.1bp \d@mx=\drawingwd \divide\d@mx by\d@my% BUG! for large \drawingwd
\includegraphics{##1\space}}}%
\def\psforoztex{%        For the OzTeX shareware on the Macintosh
%--------------
\def\PSspeci@l##1##2{%
\special{##1 \space
      ##2 1000 div dup scale
      \number-\psllx\space \number-\pslly\space translate
}}}%
\def\psfordvitps{%       From the UNIX TeXPS package, vers.>3.12
%---------------
% Convert a dimension into the number \psn@sp (in scaled points)
\def\psdimt@n@sp##1{\d@mx=##1\relax\edef\psn@sp{\number\d@mx}}
\def\PSspeci@l##1##2{%
% psfig.psr contains the def of "startTexFig": if you can locate it
% and include the correct pathname, it should work
\special{dvitps: Include0 "psfig.psr"}% contains def of "startTexFig"
\psdimt@n@sp{\drawingwd}
\special{dvitps: Literal "\psn@sp\space"}
\psdimt@n@sp{\drawinght}
\special{dvitps: Literal "\psn@sp\space"}
\psdimt@n@sp{\psllx bp}
\special{dvitps: Literal "\psn@sp\space"}
\psdimt@n@sp{\pslly bp}
\special{dvitps: Literal "\psn@sp\space"}
\psdimt@n@sp{\psurx bp}
\special{dvitps: Literal "\psn@sp\space"}
\psdimt@n@sp{\psury bp}
\special{dvitps: Literal "\psn@sp\space startTexFig\space"}
\special{dvitps: Include1 "##1"}
\special{dvitps: Literal "endTexFig\space"}
}}%
\def\psfordvialw{%   Try for dvialw, a UNIX public domain
%---------------
\def\PSspeci@l##1##2{
\special{language "PostScript",
position = "bottom left",
literal "  \psllx\space \pslly\space translate
  ##2 1000 div dup scale
  -\psllx\space -\pslly\space translate",
include "##1"}
}}%
\def\psforptips{%   For MS-DOS; LUOMA@brandeis.bitnet
%---------------
\def\PSspeci@l##1##2{{
\d@mx=\psurx bp
\advance \d@mx by -\psllx bp
\divide \d@mx by 1000\multiply\d@mx by \xscale
\incm{\d@mx}
\let\tmpx\dimincm
\d@my=\psury bp
\advance \d@my by -\pslly bp
\divide \d@my by 1000\multiply\d@my by \xscale
\incm{\d@my}
\let\tmpy\dimincm
\d@mx=-\psllx bp
\divide \d@mx by 1000\multiply\d@mx by \xscale
\d@my=-\pslly bp
\divide \d@my by 1000\multiply\d@my by \xscale
\at(\d@mx;\d@my){\special{ps:##1 x=\tmpx, y=\tmpy}}
}}}%
\def\psonlyboxes{%     Draft-like behaviour if none of the others works
%---------------
\def\PSspeci@l##1##2{%
\at(0cm;0cm){\boxit{\vbox to\drawinght
  {\vss\hbox to\drawingwd{\at(0cm;0cm){\hbox{({\tt##1})}}\hss}}}}
}}%
\def\psloc@lerr#1{%
\let\savedPSspeci@l=\PSspeci@l%
\def\PSspeci@l##1##2{%
\at(0cm;0cm){\boxit{\vbox to\drawinght
  {\vss\hbox to\drawingwd{\at(0cm;0cm){\hbox{({\tt##1}) #1}}\hss}}}}
\let\PSspeci@l=\savedPSspeci@l% restore normal output for other figs!
}}%
%\def\psfor...  add your own!
%
% Some common defs
%
\newread\pst@mpin
\newdimen\drawinght\newdimen\drawingwd
\newdimen\psxoffset\newdimen\psyoffset
\newbox\drawingBox
\newcount\xscale \newcount\yscale \newdimen\pscm\pscm=1cm
\newdimen\d@mx \newdimen\d@my
\newdimen\pswdincr \newdimen\pshtincr
\let\ps@nnotation=\relax
{\catcode`\|=0 |catcode`|\=12 |catcode`|%=12 |catcode`~=12
|catcode`#=12 |catcode`*=14
|xdef|backslashother{\}*
|xdef|percentother{%}*
|xdef|tildeother{~}*
|xdef|sharpother{#}*
}%
% useful to display special chars in \tt; fails for \,#,%
\def\R@moveMeaningHeader#1:->{}%
\def\uncatcode#1{%
\edef#1{\expandafter\R@moveMeaningHeader\meaning#1}}%
\def\execute#1{#1}% NOT stupid: cs in #1 are then identified BEFORE execution
\def\psm@keother#1{\catcode`#112\relax}% borrowed from latex
\def\executeinspecs#1{%
\execute{\begingroup\let\do\psm@keother\dospecials\catcode`\^^M=9#1\endgroup}}%
\def\@mpty{}%
% \if\matchin#1#2<=> \iftrue if #1 contains #2, <=>\iffalse otherwise:
% \if\matchexpin: idem, but #1 & #2 are first fully expanded (no \if
% inside!)
% \tmpa & \tmpb contain what's before and after the occurence of #2
\def\matchexpin#1#2{
  \fi%
%\message{(#1>#2)}
  \edef\tmpb{{#2}}%
  \expandafter\makem@tchtmp\tmpb%
  \edef\tmpa{#1}\edef\tmpb{#2}%
  \expandafter\expandafter\expandafter\m@tchtmp\expandafter\tmpa\tmpb\endm@tch%
  \if\match%
}%
\def\matchin#1#2{%
  \fi%
  \makem@tchtmp{#2}%
  \m@tchtmp#1#2\endm@tch%
  \if\match%
}%
\def\makem@tchtmp#1{\def\m@tchtmp##1#1##2\endm@tch{%
  \def\tmpa{##1}\def\tmpb{##2}\let\m@tchtmp=\relax%
  \ifx\tmpb\@mpty\def\match{YN}%
  \else\def\match{YY}\fi%
}}%
% converts any dimen in cm, with 1E-4 cm precision
\def\incm#1{{\psxoffset=1cm\d@my=#1
 \d@mx=\d@my
  \divide\d@mx by \psxoffset
  \xdef\dimincm{\number\d@mx.}
  \advance\d@my by -\number\d@mx cm
  \multiply\d@my by 100
 \d@mx=\d@my
  \divide\d@mx by \psxoffset
  \edef\dimincm{\dimincm\number\d@mx}
  \advance\d@my by -\number\d@mx cm
  \multiply\d@my by 100
 \d@mx=\d@my
  \divide\d@mx by \psxoffset
  \xdef\dimincm{\dimincm\number\d@mx}
}}%
%
%  \ReadPSize{PSfilename} reads the dimensions of a PostScript drawing
%      and stores it in \drawinght(wd)
\newif\ifNotB@undingBox
\newhelp\PShelp{Proceed: you'll have a 5cm square blank box instead of
your graphics (Jean Orloff).}%
\def\s@tsize#1 #2 #3 #4\@ndsize{
  \def\psllx{#1}\def\pslly{#2}%
  \def\psurx{#3}\def\psury{#4}%  needed by a crazyness of dvips!
  \ifx\psurx\@mpty\NotB@undingBoxtrue% this is not a valid one!
  \else
    \drawinght=#4bp\advance\drawinght by-#2bp
    \drawingwd=#3bp\advance\drawingwd by-#1bp
%  !Units related by crazy factors as bp/pt=72.27/72 should be BANNED!
  \fi
  }%
\def\sc@nBBline#1:#2\@ndBBline{\edef\p@rameter{#1}\edef\v@lue{#2}}%
\def\g@bblefirstblank#1#2:{\ifx#1 \else#1\fi#2}%
{\catcode`\%=12
\xdef\B@undingBox{%%BoundingBox}}%
%% is not a true comment in PostScript, even if % is!
\def\ReadPSize#1{
 \readfilename#1\relax
 \let\PSfilename=\lastreadfilename
 \openin\pst@mpin=#1\relax
 \ifeof\pst@mpin \errhelp=\PShelp
   \errmessage{I haven't found your postscript file (\PSfilename)}%
   \psloc@lerr{was not found}%
   \s@tsize 0 0 142 142\@ndsize
   \closein\pst@mpin
 \else
% each entry in \GlobalInputList should be unique
   \if\matchexpin{\GlobalInputList}{, \lastreadfilename}%
   \else\xdef\GlobalInputList{\GlobalInputList, \lastreadfilename}%
     \immediate\write\psbj@inaux{\lastreadfilename,}%
   \fi%
   \loop
     \executeinspecs{\catcode`\ =10\global\read\pst@mpin to\n@xtline}%
     \ifeof\pst@mpin
       \errhelp=\PShelp
       \errmessage{(\PSfilename) is not an Encapsulated PostScript File:
           I could not find any \B@undingBox: line.}%
       \edef\v@lue{0 0 142 142:}%
       \psloc@lerr{is not an EPSFile}%
       \NotB@undingBoxfalse
     \else
       \expandafter\sc@nBBline\n@xtline:\@ndBBline
       \ifx\p@rameter\B@undingBox\NotB@undingBoxfalse
         \edef\t@mp{%
           \expandafter\g@bblefirstblank\v@lue\space\space\space}%
         \expandafter\s@tsize\t@mp\@ndsize
       \else\NotB@undingBoxtrue
       \fi
     \fi
   \ifNotB@undingBox\repeat
   \closein\pst@mpin
 \fi
\message{#1}%
}%
%
% \psboxto(xdim;ydim){psfilename}: you specify the dimensions and
%    TeX uniformly scales to fit the largest one. If xdim=0pt, the
%    scale is fully determined by ydim and vice versa.
%    Notice: psboxes are a real vboxes; couldn't take hbox otherwise all
%    indentation and all cr's would be interpreted as spaces (hugh!).
%
\def\psboxto(#1;#2)#3{\vbox{%
   \ReadPSize{#3}%
   \advance\pswdincr by \drawingwd
   \advance\pshtincr by \drawinght
   \divide\pswdincr by 1000
   \divide\pshtincr by 1000
   \d@mx=#1
   \ifdim\d@mx=0pt\xscale=1000
         \else \xscale=\d@mx \divide \xscale by \pswdincr\fi
   \d@my=#2
   \ifdim\d@my=0pt\yscale=1000
         \else \yscale=\d@my \divide \yscale by \pshtincr\fi
   \ifnum\yscale=1000
         \else\ifnum\xscale=1000\xscale=\yscale
                    \else\ifnum\yscale<\xscale\xscale=\yscale\fi
              \fi
   \fi
   \divide\drawingwd by1000 \multiply\drawingwd by\xscale
   \divide\drawinght by1000 \multiply\drawinght by\xscale
   \divide\psxoffset by1000 \multiply\psxoffset by\xscale
   \divide\psyoffset by1000 \multiply\psyoffset by\xscale
   \global\divide\pscm by 1000
   \global\multiply\pscm by\xscale
   \multiply\pswdincr by\xscale \multiply\pshtincr by\xscale
   \ifdim\d@mx=0pt\d@mx=\pswdincr\fi
   \ifdim\d@my=0pt\d@my=\pshtincr\fi
   \message{scaled \the\xscale}%
 \hbox to\d@mx{\hss\vbox to\d@my{\vss
   \global\setbox\drawingBox=\hbox to 0pt{\kern\psxoffset\vbox to 0pt{%
      \kern-\psyoffset
      \PSspeci@l{\PSfilename}{\the\xscale}%
      \vss}\hss\ps@nnotation}%
   \global\wd\drawingBox=\the\pswdincr
   \global\ht\drawingBox=\the\pshtincr
   \global\drawingwd=\pswdincr
   \global\drawinght=\pshtincr
   \baselineskip=0pt
   \copy\drawingBox
 \vss}\hss}%
  \global\psxoffset=0pt
  \global\psyoffset=0pt
  \global\pswdincr=0pt
  \global\pshtincr=0pt % These are local to one figure
  \global\pscm=1cm %should not be necessary
}}%
%
% \psboxscaled{scalefactor*1000}{PSfilename} allows to bypass the
%   rounding errors of TeX integer divisions for situations where the
%   TeX box should fit the original BoundingBox with a precision
%   better
%   than 1/1000.
%
\def\psboxscaled#1#2{\vbox{%
  \ReadPSize{#2}%
  \xscale=#1
  \message{scaled \the\xscale}%
  \divide\pswdincr by 1000 \multiply\pswdincr by \xscale
  \divide\pshtincr by 1000 \multiply\pshtincr by \xscale
  \divide\psxoffset by1000 \multiply\psxoffset by\xscale
  \divide\psyoffset by1000 \multiply\psyoffset by\xscale
  \divide\drawingwd by1000 \multiply\drawingwd by\xscale
  \divide\drawinght by1000 \multiply\drawinght by\xscale
  \global\divide\pscm by 1000
  \global\multiply\pscm by\xscale
  \global\setbox\drawingBox=\hbox to 0pt{\kern\psxoffset\vbox to 0pt{%
     \kern-\psyoffset
     \PSspeci@l{\PSfilename}{\the\xscale}%
     \vss}\hss\ps@nnotation}%
  \advance\pswdincr by \drawingwd
  \advance\pshtincr by \drawinght
  \global\wd\drawingBox=\the\pswdincr
  \global\ht\drawingBox=\the\pshtincr
  \global\drawingwd=\pswdincr
  \global\drawinght=\pshtincr
  \baselineskip=0pt
  \copy\drawingBox
  \global\psxoffset=0pt
  \global\psyoffset=0pt
  \global\pswdincr=0pt
  \global\pshtincr=0pt % These are local to one figure
  \global\pscm=1cm
}}%
%
%  \psbox{PSfilename} makes a TeX box having the minimal size to
%      enclose the picture
\def\psbox#1{\psboxscaled{1000}{#1}}%
%------------------------------------------------------
%  \joinfiles file1, file2, ...n \into joinedfilename .
%     makes one file out of many
%  \splitfile joinedfilename
%     the opposite
\newif\ifn@teof\n@teoftrue
\newif\ifc@ntrolline
\newif\ifmatch
\newread\j@insplitin
\newwrite\j@insplitout
\newwrite\psbj@inaux
\immediate\openout\psbj@inaux=psbjoin.aux
\immediate\write\psbj@inaux{\string\joinfiles}%
\immediate\write\psbj@inaux{\jobname,}%
%
% INPUT REDEFINITION
%
% works if #1 is a single character
\def\toother#1{\ifcat\relax#1\else\expandafter%
  \toother@ux\meaning#1\endtoother@ux\fi}%
\def\toother@ux#1 #2#3\endtoother@ux{\def\tmp{#3}%
  \ifx\tmp\@mpty\def\tmp{#2}\let\next=\relax%
  \else\def\next{\toother@ux#2#3\endtoother@ux}\fi%
\next}%
%
% \readfilename defs:
%
\let\readfilenamehook=\relax
\def\re@d{\expandafter\re@daux}% spares typing 10 \expandafter's...
\def\re@daux{\futurelet\nextchar\stopre@dtest}%
\def\re@dnext{\xdef\lastreadfilename{\lastreadfilename\nextchar}%
  \afterassignment\re@d\let\nextchar}%
\def\stopre@d{\egroup\readfilenamehook}%
\def\stopre@dtest{%
  \ifcat\nextchar\relax\let\nextread\stopre@d
  \else
    \ifcat\nextchar\space\def\nextread{%
      \afterassignment\stopre@d\chardef\nextchar=`}%
    \else\let\nextread=\re@dnext
      \toother\nextchar
      \edef\nextchar{\tmp}%
    \fi
  \fi\nextread}%
\def\readfilename{\bgroup%
  \let\\=\backslashother \let\%=\percentother \let\~=\tildeother
  \let\#=\sharpother \xdef\lastreadfilename{}%
  \re@d}%
%
% redefines \input using \readfilename
%
\xdef\GlobalInputList{\jobname}%
\def\psnewinput{%
  \def\readfilenamehook{% each entry in \GlobalInputList should be unique
    \if\matchexpin{\GlobalInputList}{, \lastreadfilename}%
    \else\xdef\GlobalInputList{\GlobalInputList, \lastreadfilename}%
      \immediate\write\psbj@inaux{\lastreadfilename,}%
    \fi%
    \ps@ldinput\lastreadfilename\relax%
    \let\readfilenamehook=\relax%
  }\readfilename%
}%
\expandafter\ifx\csname @@input\endcsname\relax    % then Plain
  \immediate\let\ps@ldinput=\input\def\input{\psnewinput}%
\else
  \immediate\let\ps@ldinput=\@@input
  \def\@@input{\psnewinput}%
\fi%
\def\nowarnopenout{%
 \def\warnopenout##1##2{%
   \readfilename##2\relax
   \message{\lastreadfilename}%
   \immediate\openout##1=\lastreadfilename\relax}}%
\def\warnopenout#1#2{%
 \readfilename#2\relax
 \def\t@mp{TrashMe,psbjoin.aux,psbjoint.tex,}\uncatcode\t@mp
 \if\matchexpin{\t@mp}{\lastreadfilename,}%
 \else
   \immediate\openin\pst@mpin=\lastreadfilename\relax
   \ifeof\pst@mpin
     \else
     \errhelp{If the content of this file is so precious to you, abort (ie
press x or e) and rename it before retrying.}%
     \errmessage{I'm just about to replace your file named \lastreadfilename}%
   \fi
   \immediate\closein\pst@mpin
 \fi
 \message{\lastreadfilename}%
 \immediate\openout#1=\lastreadfilename\relax}%
% % will have an unusual catcode below; use * instead
%\vbox
{\catcode`\%=12\catcode`\*=14
\gdef\splitfile#1{*
 \readfilename#1\relax
 \immediate\openin\j@insplitin=\lastreadfilename\relax
 \ifeof\j@insplitin
   \message{! I couldn't find and split \lastreadfilename!}*
 \else
   \immediate\openout\j@insplitout=TrashMe
   \message{< Splitting \lastreadfilename\space into}*
   \loop
     \ifeof\j@insplitin
       \immediate\closein\j@insplitin\n@teoffalse
     \else
       \n@teoftrue
       \executeinspecs{\global\read\j@insplitin to\spl@tinline\expandafter
         \ch@ckbeginnewfile\spl@tinline%Beginning-Of-File-Named:%\endcheck}*
       \ifc@ntrolline
       \else
         \toks0=\expandafter{\spl@tinline}*
         \immediate\write\j@insplitout{\the\toks0}*
       \fi
     \fi
   \ifn@teof\repeat
   \immediate\closeout\j@insplitout
 \fi\message{>}*
}*
\gdef\ch@ckbeginnewfile#1%Beginning-Of-File-Named:#2%#3\endcheck{*
 \def\t@mp{#1}*
 \ifx\@mpty\t@mp
   \def\t@mp{#3}*
   \ifx\@mpty\t@mp
     \global\c@ntrollinefalse
   \else
     \immediate\closeout\j@insplitout
     \warnopenout\j@insplitout{#2}*
     \global\c@ntrollinetrue
   \fi
 \else
   \global\c@ntrollinefalse
 \fi}*
\gdef\joinfiles#1\into#2{*
 \message{< Joining following files into}*
 \warnopenout\j@insplitout{#2}*
 \message{:}*
 {*
 \edef\w@##1{\immediate\write\j@insplitout{##1}}*
\w@{% This collection of files was produced with CERN psbox package}*
\w@{% To decompose and tex it:}*
\w@{%-save this with a filename CONTAINING ONLY LETTERS and a .TEX}*
\w@{% extension (say, JOINTFIL.TEX), in some uncrowded directory;}*
\w@{%-make sure you can \string\input\space psbox.tex (version>=1.3);}*
\w@{%  (else ftp cs.nyu.edu(=128.122.140.24):pub/TeX/psbox/, then get}*
\w@{%  and tex the file psboxall.tex; more info in psbREAD.ME)}*
\w@{%-tex JOINTFIL.TEX using Plain, or LaTeX, or whatever is needed by}*
\w@{%  the first file in the joining (after splitting JOINTFIL.TEX into}*
\w@{%  it's constituents, TeX will try to process it as it stands).}*
\w@{\string\input\space psbox.tex}*
\w@{\string\splitfile{\string\jobname}}*
\w@{\string\let\string\autojoin=\string\relax}*
}*
 \expandafter\tre@tfilelist#1, \endtre@t
 \immediate\closeout\j@insplitout
 \message{>}*
}*
\gdef\tre@tfilelist#1, #2\endtre@t{*
 \readfilename#1\relax
 \ifx\@mpty\lastreadfilename
 \else
   \immediate\openin\j@insplitin=\lastreadfilename\relax
   \ifeof\j@insplitin
     \errmessage{I couldn't find file \lastreadfilename}*
   \else
     \message{\lastreadfilename}*
     \immediate\write\j@insplitout{%Beginning-Of-File-Named:\lastreadfilename}*
     \executeinspecs{\global\read\j@insplitin to\oldj@ininline}*
     \loop
       \ifeof\j@insplitin\immediate\closein\j@insplitin\n@teoffalse
       \else\n@teoftrue
         \executeinspecs{\global\read\j@insplitin to\j@ininline}*
         \toks0=\expandafter{\oldj@ininline}*
         \let\oldj@ininline=\j@ininline
         \immediate\write\j@insplitout{\the\toks0}*
       \fi
     \ifn@teof
     \repeat
   \immediate\closein\j@insplitin
   \fi
   \tre@tfilelist#2, \endtre@t
 \fi}*
}%
% To be put at the end of a file, for making a tar-like file containing
%   everything it used.
\def\autojoin{%
 \immediate\write\psbj@inaux{\string\into{psbjoint.tex}}%
 \immediate\closeout\psbj@inaux
 \expandafter\joinfiles\GlobalInputList\into{psbjoint.tex}%
}%
%----------------------------------------------------------------
%  Annotations & Captions etc...
%
%
% \centinsert{anybox} is just a centered \midinsert, but is included as
%    people barely use the original inserts from TeX.
%
\def\centinsert#1{\midinsert\line{\hss#1\hss}\endinsert}%
\def\psannotate#1#2{\vbox{%
  \def\ps@nnotation{#2\global\let\ps@nnotation=\relax}#1}}%
\def\pscaption#1#2{\vbox{%
   \setbox\drawingBox=#1
   \copy\drawingBox
   \vskip\baselineskip
   \vbox{\hsize=\wd\drawingBox\setbox0=\hbox{#2}%
     \ifdim\wd0>\hsize
       \noindent\unhbox0\tolerance=5000
    \else\centerline{\box0}%
    \fi
}}}%
% for compatibility with older versions, but \psfig is a bad name!
%\def\psfig#1#2#3{\pscaption{\psannotate{#1}{#2}}{#3}}
%\def\psfigurebox#1#2#3{\pscaption{\psannotate{\psbox{#1}}{#2}}{#3}}
%
% \at(#1;#2)#3 puts #3 at #1-higher and #2-right of the current
%    position without moving it (to be used in annotations).
\def\at(#1;#2)#3{\setbox0=\hbox{#3}\ht0=0pt\dp0=0pt
  \rlap{\kern#1\vbox to0pt{\kern-#2\box0\vss}}}%
%
% \gridfill(ht;wd) makes a 1cm*1cm grid of ht by wd whose lower-left
%   corner is the current point
\newdimen\gridht \newdimen\gridwd
\def\gridfill(#1;#2){%
  \setbox0=\hbox to 1\pscm
  {\vrule height1\pscm width.4pt\leaders\hrule\hfill}%
  \gridht=#1
  \divide\gridht by \ht0
  \multiply\gridht by \ht0
  \gridwd=#2
  \divide\gridwd by \wd0
  \multiply\gridwd by \wd0
  \advance \gridwd by \wd0
  \vbox to \gridht{\leaders\hbox to\gridwd{\leaders\box0\hfill}\vfill}}%
%
% Useful to measure where to put annotations
\def\fillinggrid{\at(0cm;0cm){\vbox{%
  \gridfill(\drawinght;\drawingwd)}}}%
%
% \textleftof\anybox: Sample text\endtext
%   inserts "Sample text" on the left of \anybox ie \vbox, \psbox.
%   \textrightof is the symmetric (not documented, too uggly)
% Welcome any suggestion about clean wraparound macros from
%   TeXhackers reading this
%
\def\textleftof#1:{%
  \setbox1=#1
  \setbox0=\vbox\bgroup
    \advance\hsize by -\wd1 \advance\hsize by -2em}%
\def\textrightof#1:{%
  \setbox0=#1
  \setbox1=\vbox\bgroup
    \advance\hsize by -\wd0 \advance\hsize by -2em}%
\def\endtext{%
  \egroup
  \hbox to \hsize{\valign{\vfil##\vfil\cr%
\box0\cr%
\noalign{\hss}\box1\cr}}}%
%
% \frameit{\thick}{\skip}{\anybox}
%    draws with thickness \thick a box around \anybox, leaving \skip of
%    blank around it. eg \frameit{0.5pt}{1pt}{\hbox{hello}}
% \boxit{\anybox} is a shortcut.
\def\frameit#1#2#3{\hbox{\vrule width#1\vbox{%
  \hrule height#1\vskip#2\hbox{\hskip#2\vbox{#3}\hskip#2}%
        \vskip#2\hrule height#1}\vrule width#1}}%
\def\boxit#1{\frameit{0.4pt}{0pt}{#1}}%
\catcode`\@=12 % cs containing @ are unreachable
%
% CUSTOMIZE YOUR DEFAULT DRIVER:
%    Uncomment the line corresponding to your TeX system:
%\psfortextures%     For TeXtures on the Macintosh
%\psforoztex   %     For OzTeX shareware on the Macintosh
%\psfordvitops %     For the DVItoPS converter for TeX on IBM mainframes
\psfordvips   %     For DVIPS converter on VAX and UNIX
%\psfordvitps  %     For dvitps from TeXPS package under UNIX
%\psfordvialw  %     For dvialw, UNIX public domain
%\psonlyboxes  %     Blank Boxes (when all else fails).